\documentclass[aps,prd,nofootinbib,superscriptaddress,twocolumn]{revtex4}
\usepackage{hyperref}
\usepackage{amsfonts}

\usepackage{amsmath}
\usepackage{graphicx}
\usepackage{color}
\usepackage{mathtools,leftidx}

\DeclareMathAlphabet\mathbfcal{OMS}{cmsy}{b}{n}

\begin{document}

\title{Optical response of a topological-insulator--quantum-dot hybrid interacting with a probe electric field}

\author{L. A. Castro-Enriquez}
\email{lacastro@uniquindio.edu.co}
\affiliation{Programa de F\'{i}sica, Universidad del Quind\'{i}o, 630001 Armenia, Colombia.}

\author{L. F. Quezada}
\email{lfqm1987@ciencias.unam.mx}
\affiliation{Centro de Innovaci\'{o}n y Desarrollo Tecnol\'{o}gico en C\'{o}mputo, Instituto Polit\'{e}cnico Nacional, UPALM, 07700 Ciudad de M\'{e}xico, M\'{e}xico}

\author{A. Mart\'{i}n-Ruiz}
\email{alberto.martin@nucleares.unam.mx}
\affiliation{Instituto de Ciencias Nucleares, Universidad Nacional Aut\'{o}noma de M\'{e}xico, 04510 Ciudad de M\'{e}xico, M\'{e}xico}

\begin{abstract}
\noindent
We study the interaction between a topological insulator nanoparticle and a quantum dot subject to an applied electric field. The electromagnetic response of the topological insulator is derived from axion electrodynamics in the quasistatic approximation. Localized modes are quantized in terms of dipolar bosonic modes, which couples dipolarly to the quantum dot. Hence, we treat the hybrid as a two-level system interacting with a single bosonic mode, where the coupling strength encodes the information concerning the nontrivial topology of the nanoparticle. The interaction of the hybrid with the environment is implemented through the coupling with a continuum reservoir of radiative output modes and a reservoir of phonon modes. In particular, we use the method of Zubarev's Green functions to derive an expression for the optical absorption spectrum of the system. We apply our results to a realistic system which consists of a topological insulator nanoparticle made of TlBiSe$_{2}$ interacting with a cadmium selenide quantum dot, both immersed in a polymer layer such as poly(methyl methacrylate). The optical absorption spectrum exhibits Fano resonances with a line shape that strongly depends on the polarization of the electric field as well as on the topological magnetoelectric polarizability $\theta$. Our results and methods can also be applied to nontopological magnetoelectric materials such as Cr$_{2}$O$_{3}$.
\end{abstract}

\maketitle

\section{Introduction} \label{intro}

Recently, topological insulators (TI) have been investigated intensively both theoretically and experimentally \cite{TI-Hasan, TI-Qi}. These materials are fully gapped in the bulk, but have gapless edge or surface states which are topologically protected by time-reversal (TR) symmetry. The surface states of a three-dimensional (3D) TI consist of an odd number of massless Dirac cones, whose existence is ensured by the $\mathbb{Z} _{2}$ topological invariant of the bulk \cite{Fu, Moore}. Furthermore, Kramers theorem guarantees that no TR invariant perturbation can open up an insulating gap at the Dirac point on the surface. However, a TI becomes a fully gapped system (both in the bulk and on the surface) if a TR breaking perturbation is introduced on the surface. In this case, the electromagnetic response of a 3D TI is described by the topological $\theta$ term of the form \cite{Qi-TFT}
\begin{align}
S _{\theta} = \frac{\alpha}{\pi} \sqrt{\frac{\epsilon _{0}}{\mu _{0}}} \int d ^{4} x \, \theta \, \vec{E} \cdot \vec{B} , \label{Action}
\end{align}
where $\vec{E}$ and $\vec{B}$ are the electromagnetic fields, $\alpha = e ^{2}/ 2 \epsilon _{0} h c \approx 1/ 137$ is the fine structure constant, and $\theta$ is the topological magnetoelectric polarization. $\theta = 0$ describes a conventional insulator, whereas $\theta = \pi$ describe topological insulators. Such a physically measurable and topologically nontrivial response originates from the Dirac fermions on the surface of the TI.

On the other hand, the optical properties of hybrid systems composed by semiconductor quantum dots (QDs) and plasmonic nanostructures (such as spherical metallic nanoparticles and metallic nanorods), have attracted great attention because of the possible broad range of applications in photonics and optoelectronics. When these components are close enough, the interaction between excitons from the QD and the surface plasmons significantly influences the optical properties of the system and leads to several interesting phenomena, such as Fano resonances \cite{Fano, Fano2, Miroshnichenko} and plasmonic meta-resonances \cite{MetaReso, MetaReso2}. To date, there have been several quantum and semiclassical studies of the interaction between dipole emitters and metallic nanoparticles \cite{Waks, Manjavacas, Alpeggiani, Kosionis, Artuso, RCGe, Ahmad, Salmonogli, Naeimi}.

Recent advances in the fabrication of nanostructured devices made from topological insulator materials, such as TI nanoparticles \cite{Cho, Kershaw, Vargas, Jia, Claro, Rider} and TI nanowires \cite{Peng, Xiu, Dufouleur, Hong, Dellabetta, Jauregui, Siroki}, mark a step towards utilizing topological properties at the nanoscale in applications such as quantum computing, photonics and optoelectronics. Also, they provide an additional scenario where the topological magnetoelectric effect, as described by the action (\ref{Action}), can be tested. This is precisely the main motivation of this work, where we pursued the idea that quantum emitters near to topological insulator nanostructures could shed information on the topological nontriviality of the materials.

In this paper we study the response of a hybrid nanostructure consisting of a quantum dot (QD) coupled to a topological insulator nanoparticle (TINP), subject to an applied electric field. The field couples to both the QD and the TINP, and all three constituents interact with each other through a dipole-dipole coupling. The electromagnetic field of the TINP is derived within the quasistatic approximation, and we show that it supports well-defined dipolar bosonic modes. In a realistic model one has to consider the finite lifetime of the excitations of the system, which produce finite widths in the corresponding spectral resonances. In general, such widths are the result of the inelastic interaction with a continuum of modes. Here we describe this inelastic interactions by coupling the system with a continuum reservoir of radiative output modes and a reservoir of phonon modes. Taking into account all the interactions, we use the method of Zubarev's Green functions to calculate the absorption spectrum of the system. For numerical calculations we consider the specific case of a Cadmium Selenide (CdSe) QD in proximity to a TINP made of TlBiSe$_{2}$.

\begin{figure}
\includegraphics[scale=0.35]{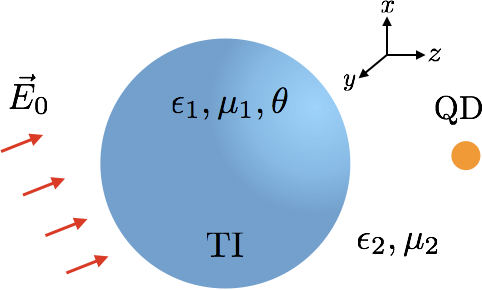}
\caption{\small Schematic of the topological insulator-quantum dot hybrid in the presence of an external electric field. We refer longitudinal (transverse) coupling to the configuration where the electric field polarization points along (perpendicular) to the line connecting the center of the TI and the QD position.}
\label{TI-QD-Fig}
\end{figure}

The paper is organized as follows. In Sec. \ref{electroTIs} we compute the electromagnetic response of a spherical topological insulator nanoparticle interacting with a probe electric field, as depicted in fig. \ref{TI-QD-Fig}. Then, in Sec. \ref{QuantumOpticalModel}, we turn to the quantization of the electromagnetic field modes on the TI surface, from which we derive a quantum-optical model to describe the interaction between a TI nanoparticle and a quantum-dot nearby. Furthermore, we include damping effects due to the interaction of the system with a continuum reservoir of radiative output modes and a reservoir of phonon modes. In Sec. \ref{AbsSpectrumSect} we use the method of Zubarev's Green function to compute the optical absorption spectrum of the TI-QD hybrid. We apply our results to a system in which a TI nanoparticle made of TlBiSe$_{2}$ interacts with a Cadmium Selenide (CdSe) QD. Finally in Sec. \ref{ConclusionSection} we summarize the main results of the paper.

\section{Electromagnetic field distribution in the quasistatic approximation} \label{electroTIs}

In this paper we are concerned with the problem of a TI-QD hybrid interacting with a probe electric field. To move toward this goal, in this section we investigate the electrodynamics of the problem. After a brief review of the electromagnetic response of TIs, we calculate the electromagnetic field distribution due to an spherical topological insulator in a monochromatic electric field, and with the aim of quantizing the TI field, we finally obtain localized solutions.

\subsection{Electrodynamics of topological insulators}

The electromagnetic response of a system in the presence of the $\theta$ term (\ref{Action}) is still described by the ordinary Maxwell equations but with the modified constitutive equations (in SI units) \cite{Qi-TFT}
\begin{align}
\vec{D} &= \epsilon  \epsilon _{0} \, \vec{E} + \alpha (\theta / \pi) \sqrt{\frac{\epsilon _{0}}{\mu _{0}}} \vec{B} , \label{ConstRelD} \\ \vec{H} &= \frac{\vec{B}}{\mu \mu _{0}} - \alpha (\theta / \pi) \sqrt{\frac{\epsilon _{0}}{\mu _{0}}} \vec{E} , \label{ConstRelH}
\end{align}
where $\epsilon$ and $\mu$ are the relative permittivity and permeability of the medium. The corresponding vacuum quantities are $\epsilon _{0} = 8.85 \times 10 ^{-12}$ F/m and $\mu _{0} = 4 \pi \times 10 ^{-7}$ H/m. The description of a TI in terms of the modified constitutive relations incorporating the topological magnetoelectric effect is only valid when the massless topological surfaces modes are gapped. In the case of nonmagnetic topological insulators, this is achieved by means of a magnetic perturbation (applied field and/or film coating) \cite{Burkov}. Recent advances in experimental condensed matter physics has allow the growth of intrinsic magnetic topological insulators, that is, ones that have magnetic properties by their own since they are composed by atoms with spinful nuclei \cite{MagneticTI1, MagneticTI2, MagneticTI3}. Interestingly, despite this magnetic order at atomic scale, the whole material lacks of bulk magnetization, however it works as a source of intrinsic TR breaking perturbation that gap the surface states \cite{ExpMagneticTI, ExpMagneticTI2}. Hence, axion electrodynamics as described above works for both magnetic and nonmagnetic topological insulators \cite{MagneticTI1, MagneticTI2, MagneticTI3}. Once the surface Dirac fermions are gapped, $\theta$ is quantized in odd integer values of $\pi$ such that $\theta = (2n + 1) \pi$, where the value of $n$ is determined by the nature of the TR-breaking perturbation. In this work we consider that the TR perturbation is a magnetic coating of small thickness whose magnetization points outward the TI. Indeed, it corresponds to modifying the interface by adsorbing surface layers of nonzero Chern number \cite{Essin}. 

As mentioned above, nontrivial effects due to the topological $\theta$ term appear only at the interface $\Sigma$ of a TI in contact with a trivial insulator (or vacuum), where the TMEP suddenly changes. Assuming that the time derivatives of the fields are finite in the vicinity of $\Sigma$, Maxwell equations imply boundary conditions, which, for vanishing free sources on $\Sigma$, read
\begin{align}
[ \vec{n} \cdot ( \epsilon \vec{E}) ] _{\Sigma} = \tilde{\alpha} c \, \vec{n} \cdot \vec{B} \vert _{\Sigma} , \quad [ \vec{n} \times \vec{E} ] _{\Sigma} = \vec{0} , \notag \\[5pt] [ \vec{n} \cdot \vec{B} ] _{\Sigma} = 0 , \quad [ \vec{n} \times ( \vec{B} / \mu ) ] _{\Sigma} = ( \tilde{\alpha} / c ) \, \vec{E} \times \vec{n} \vert _{\Sigma} , \label{BoundaryConditions}
\end{align}
where $\tilde{\alpha} = \alpha (\theta / \pi)$, $c = ( \mu _{0} \epsilon _{0} ) ^{-1/2}$ is the speed of light in vacuum, and $\vec{n}$ is the outward unit normal to $\Sigma$. Further, the notation is $[\vec{F}] _{\Sigma} = \vec{F} _{\mbox{\scriptsize out}} - \vec{F} _{\mbox{\scriptsize in}}$, where the subscript ``in" (``out") refers to inside (outside) the TI. It is worth mentioning that the boundary conditions are perfectly consistent in relating field discontinuities at the interface with components of the fields which are continuous there. A number of magnetoelectric effects have been predicted on the basis of this theory \cite{Qi-monopole, Karch, Chang, Cortijo, MCU1, Ge, Crosse, MCU2, MCU3, Crosse2, MC, Campos, MU, MR, MU2, Bonilla, Franca, Nogueira}. However, it has been experimentally verified only through the measurement of Kerr and Faraday angles at the surface of a strained HgTe 3D TI \cite{Dziom}.

\subsection{Spherical TI in a monochromatic electric field}

With the aim of quantizing the TI electromagnetic field modes, we begin with its classical electromagnetic description. Let us consider a spherical topological insulator of radius $R$ in the presence of an electric field $\vec{E} (\vec{r} , t)$, as shown in Fig. \ref{TI-QD-Fig}. The TI is characterized by a dielectric function $\epsilon _{1} (\omega )$, a permeability function $\mu _{1} (\omega)$, and a topological magnetoelectric polarizability $\theta$, while the dielectric outside the TI has dielectric function $\epsilon _{2} (\omega )$ and magnetic permeability $\mu _{2} (\omega)$. The TI size is assumed to be small compared to the wavelength of the applied electric field, so we can make the time harmonic approximation. That is, the electric and magnetic fields can be written respectively as $\vec{E} (\vec{r} , t) \! = \! \mbox{Re} \big\{\ \!\!\! \vec{E} (\vec{r} \, ) e ^{- i \omega t} \big\}\ \!\!$ and $\vec{B} (\vec{r} , t) \! = \! \mbox{Re} \big\{\ \!\!\! \vec{B} (\vec{r} \, ) e ^{- i \omega t} \big\}\ \!\!$, where $\vec{E} (\vec{r} \,)$ and $\vec{B} (\vec{r} \,)$ are the fields associated with the solution of the static Maxwell equations and $\mbox{Re} \{ \, \}$ indicates the real part. In other words, the quasistatic approximation consists of neglecting the retardation effects everywhere except in the dielectric and permeability functions dependence on the frequency. Here, we shall consider that the TI is driven by an external monochromatic field, i.e., $\vec{E} (\vec{r} , t) \! = \! \mbox{Re} \big\{\ \!\!\! \vec{E} _{0} e ^{- i \omega t} \big\}\ \!\!$, where $\vec{E} _{0}$ is a constant vector. Therefore, in order to determine the electromagnetic field distribution in the quasistatic approximation, we have to solve the problem of a spherical TI in a constant electric field. This is a simple but not straightforward task, so we leave the details to Appendix \ref{CalculationsFields}. We find that the electric field can be written as a sum of the externally applied electric field $\vec{E} _{0}$ plus the electric field of the topological insulator given by
\begin{align}
\vec{\mathcal{E}} (\vec{r}, \omega) &= \sum _{i = x,y,z} \frac{\epsilon _{2} - \epsilon _{1} - \mu _{\mbox{\scriptsize e}} \tilde{\alpha} ^{2}}{2 \epsilon _{2} + \epsilon _{1} + \mu _{\mbox{\scriptsize e}} \tilde{\alpha} ^{2}} E _{0 i} \, \vec{\mathcal{G}} _{i} (\vec{r} \, ) , \label{E-TI}
\end{align}
while the induced magnetic field, which is a pure TI response, becomes
\begin{align}
\vec{\mathcal{B}} (\vec{r}, \omega) &= \sum _{i = x,y,z} \xi (r) \frac{3 \epsilon _{2} \, \mu _{\mbox{\scriptsize e}} \, \tilde{\alpha} / 2c }{2 \epsilon _{2} + \epsilon _{1} + \mu _{\mbox{\scriptsize e}} \tilde{\alpha} ^{2}} E _{0i} \, \vec{\mathcal{G}} _{i} (\vec{r} \,) , \label{B-TI}
\end{align}
where $\xi (r) = 1$ for $r>R$ and $\xi (r) = - 2$ for $r<R$. Here, $\mu _{\mbox{\scriptsize e}} = 2 \mu _{1} \mu _{2} / (\mu _{1} + 2 \mu _{2})$, and $E _{0i}$ is the $i$th component of $\vec{E} _{0}$. In these expressions we have defined the dimensionless vector
\begin{align}
    \vec{\mathcal{G}} _{i} (\vec{r} \,) = \left\lbrace  \begin{array}{c}  \vec{e} _{i} \\[5pt] - \frac{R ^{3}}{r ^{3}} \left[ 3 (\vec{e} _{i} \cdot \vec{e} _{r}) \vec{e} _{r} - \vec{e} _{i} \right]    \end{array} \right.  \begin{array}{c} r<R \\[5pt] r>R  \end{array} , \label{G-vector}
\end{align}
where $\vec{e} _{r}$ is a unit vector pointing in the direction of $\vec{r}$, and $r$ is the center-to-center distance from the TI to the QD.

It is clear that the above fields satisfy the orthogonality relation
\begin{align}
    \int \mathcal{F} _{i} (\vec{r}, \omega) \, \mathcal{F} _{j} (\vec{r}, \omega) \, d ^{3} \vec{r}  &= 0 , \qquad i \neq j ,
\end{align}
where $\mathcal{F} _{i} = \mathcal{E} _{i} , \mathcal{B} _{i}$.

To compute the time-dependent EM fields, we have to Fourier transform the above results. In general, the permittivity and permeability functions are frequency-dependent. However, since most topological insulators are nonmagnetic in the bulk, we henceforth assume the permeabilities to be constant, such that the only frequency-dependence is through the permittivity $\epsilon _{1} (\omega)$. Let us recall that even intrinsic magnetic topological insulators has turned out to be antiferromagnetic in the bulk, and as such they have vanishing magnetization \cite{ExpMagneticTI, ExpMagneticTI2}. Thus we can safely take $\mu _{1} (\omega) \approx \mu _{1}$ and the following analysis is valid for both magnetic and nonmagnetic TIs. Hence, to obtain the time-dependent EM fields, a model for the dielectric function is necessary. Because of the low concentration of free carriers in insulators the most general phenomenological model to describe the optical response of a TI is a sum of oscillators to account for particular absorption resonances. Here, we consider a single-mode model for the dielectric function given by
\begin{align}
    \epsilon _{1} (\omega) = 1 + \frac{\omega _{e} ^{2}}{\omega _{R} ^{2} - \omega ( \omega + i \gamma _{0} )} . \label{DielectricFunction}
\end{align}
In this model, $\omega _{R}$ is the resonant frequency of the oscillator while $\omega _{e}$ accounts for the oscillator strength. The damping parameter $\gamma _{0}$ satisfying $\gamma _{0} \ll \omega _{R}$ accounts for energy dissipation due to ohmic losses in the TI. In Appendix \ref{Approximations} we show that when $\gamma _{0} \ll \omega$ the fields of the TI can be written as
\begin{align}
\vec{\mathcal{E}} (\vec{r}, \omega) & \approx \sum _{i = x,y,z} \eta \frac{\omega _{0} ^{2} / 2 \Omega}{\omega - \Omega + i \gamma _{0} / 2} E _{0 i} \, \vec{\mathcal{G}} _{i} (\vec{r} \, ) , \label{E-TI2} \\ \vec{\mathcal{B}} (\vec{r}, \omega) & \approx \sum _{i = x,y,z} \eta \, ( \mu _{\mbox{\scriptsize e}} \, \tilde{\alpha} / 2 c )  \xi (r) \frac{\omega _{0} ^{2} / 2 \Omega}{\omega - \Omega + i \gamma _{0} / 2} E _{0 i} \, \vec{\mathcal{G}} _{i} (\vec{r} \,) , \label{B-TI2}
\end{align}
where $\omega _{0} = \omega _{e} / \sqrt{2 \epsilon _{2} + 1 + \mu _{\mbox{\scriptsize e}} \tilde{\alpha} ^{2}}$, $\eta = 3 \epsilon _{2} / (2 \epsilon _{2} + 1 + \mu _{\mbox{\scriptsize e}} \tilde{\alpha} ^{2})$ and $\Omega = \sqrt{\omega _{0} ^{2} + \omega _{R} ^{2}}$. We observe that in this limit the electromagnetic fields follow Lorentzian spectra, whose approximation is appropriate when the TI is interacting with a dipole whose resonant frequency is close to plasmon resonance. This will be our main interest in this paper.

\subsection{Localized solutions}

In order to quantize the optical modes, we need to find localized solutions to the field equations. These solutions are bounded in space and decay to zero in the far field domain. The electromagnetic fields derived previously are not localized solutions because they are driven by a monochromatic plane wave that extends infinitely in space. Localized solutions are obtained by exciting the TI with an impulse function rather than by a monochromatic field, since after the impulse has ended, only the localized modes will remain. Taking an input field of the form $E _{0} (t) = E _{0} \, \delta (t) $, the electric field of the topological insulator in the time-harmonic domain is given by
\begin{align}
\vec{\mathcal{E}} (\vec{r}, t) & = \sum _{i = x,y,z} \Lambda _{i} \sin (\Omega t ) \, e ^{- \gamma _{0} t /2} \, \vec{\mathcal{G}} _{i} (\vec{r} \,) , \label{ElectricField}
\end{align}
where $\Lambda _{i} = E _{0 i} \, \eta \, (\omega _{0} ^{2} / 2 \Omega)$. The induced magnetic field is proportional to the electric field, i.e., $\vec{\mathcal{B}} (\vec{r}, t) = ( \mu _{\mbox{\scriptsize e}} \, \tilde{\alpha} / 2 c ) \, \xi (r) \, \vec{\mathcal{E}} (\vec{r}, t) $. One can readily verify that these fields represent localized solutions to the field equations: $\vec{\mathcal{G}}$ satisfies both the field equations and the boundary conditions at $\omega = \Omega$ in the undamped limit.

\section{Quantum-optical model for dipole-TI interaction} \label{QuantumOpticalModel}

\subsection{Quantization of the TI response}

In order to quantize the confined field modes on the surface of the TI, we ignore for the moment the term $\gamma_{0}$, so the fields are assumed to be a steady-state sinusoidal functions. Such term will be included later with the incorporation of a continuum of reservoir modes.

Because of the orthogonality of the three modes of the TI, we may quantize each one of them individually. To this end, we start with the energy of the $i$th mode of the EM fields:
\begin{align}
U _{i} &= \frac{\epsilon _{0}}{2} \Lambda _{i} ^{2} \sin ^{2} ( \Omega t) \int d ^{3} \vec{r} \;\, \vert \vec{\mathcal{G}} _{i} (\vec{r} \, ) \vert ^{2} \notag \\ & \phantom{=} \hspace{1cm} \times \Bigg\{\ \!\! \frac{d [ \mbox{Re}( \omega \epsilon ) ]}{d \omega} \Bigg| _{\omega = \Omega}  \! + \! \frac{\mu _{\mbox{\scriptsize e}} ^{2} \tilde{\alpha} ^{2}}{4 \mu} \xi ^{2} (r) \Bigg\}\  , \label{EnergyStored}
\end{align}
where we have used the proper definition of energy for dispersive materials. Note that the second term in the braces is a contribution from the induced magnetic field. Field quantization is followed by definition of the normalized amplitude $A _{i} = \Lambda _{i} / \mathcal{N}$, where
\begin{align}
   \frac{1}{\mathcal{N} ^{2}} \! = \! \frac{\epsilon _{0}}{2 \hbar \Omega} \! \int \!\! d ^{3} \vec{r} \; \vert \vec{\mathcal{G}} _{i} (\vec{r} \, ) \vert ^{2}   \Bigg\{\ \!\! \frac{d [ \mbox{Re}( \omega \epsilon ) ]}{d \omega} \Bigg| _{\omega = \Omega}  \!\!\! + \! \frac{\mu _{\mbox{\scriptsize e}} ^{2} \tilde{\alpha} ^{2}}{4 \mu} \xi ^{2} (r) \Bigg\}\ \! , \label{Normalization}
\end{align}
such that the energy (\ref{EnergyStored}) can be rewritten in the simplest form
\begin{align}
    U _{i} = \hbar \Omega A _{i} ^{2} \sin ^{2} ( \Omega t) . \label{EnergyStored2}
\end{align}
The normalization factor (\ref{Normalization}) is explicitly evaluated in Appendix \ref{AppNormalization}. Equation (\ref{EnergyStored2}) only gives the energy stored in the electromagnetic fields, but for energy conservation and Hamiltonian estimation, it needs to consider another form of energy due to current flowing in the TI surface. To maintain energy conservation, we add a second term in the Hamiltonian accounting for the periodic conversion between stored potential energy (represented by the energy of the field) and kinetic energy due to current flowing  in the TI. This energy must be of the form,
\begin{align}
    K _{i} = \hbar \Omega A _{i} ^{2} \cos ^{2} ( \Omega t) , \label{KineticEnergy}
\end{align}
such that the total energy $H _{i} = U _{i} + K _{i} = \hbar \Omega A _{i} ^{2}$ is constant at all times. Therefore, we suppose the topological insulator acts as a normal harmonic oscillator where the energy is periodically converted from potential energy to kinetic energy. Defining a time-dependent amplitude, $\mathcal{A} _{i} (t) = A _{i} \sin (\Omega t)$, the total Hamiltonian of the field modes can be written as
\begin{align}
    H _{i} = \frac{\hbar}{\Omega} \left( \dot{\mathcal{A}} _{i} ^{2} + \Omega ^{2} \mathcal{A} _{i} ^{2} \right) ,
\end{align}
where $\dot{\mathcal{A}} _{i}$ is the time derivative of $\mathcal{A} _{i}$. The substitution $\hbar \to m \Omega / 2$ allows us to interpret the above Hamiltonian as that of a mechanical oscillator with mass $m$ and resonant frequency $\Omega$. Therefore, the two variables $\mathcal{A} _{i}$ and $2 \hbar \dot{\mathcal{A}} _{i} / \Omega$ form a pair of canonical conjugate variables that can be quantized. So, in order to quantize the TI response, we promote these two conjugate variables to quantum operators as $\mathcal{A} _{i} \to \hat{x} _{i}$ and $2 \hbar \dot{\mathcal{A}} _{i} / \Omega \to \hat{p} _{i}$, which satisfy the commutation relation $\left[\hat{x} _{i} , \hat{p} _{j} \right] = i \delta _{ij} \hbar $. Now, introducing the bosonic creation and annihilation operators,
\begin{align}
    \hat{a} _{i} = \hat{x} _{i} + \frac{i}{2 \hbar} \hat{p} _{i} , \qquad \hat{a} _{i} ^{\dagger} = \hat{x} _{i} - \frac{i}{2 \hbar} \hat{p} _{i} ,  \label{BosonicOperators}
\end{align}
satisfying the commutation relation $[\hat{a} _{i} , \hat{a} _{j} ^{\dagger} ] = \delta _{ij}$, the definition of the Hamiltonian takes the simple form,
\begin{align}
    \hat{H} = \hbar \, \Omega \! \sum _{i = x,y,z} \left( \hat{a} ^{\dagger} _{i} \hat{a} _{i} + 1/2 \right). \label{HamOperatorTI}
\end{align}
Now we proceed with the quantization of the electromagnetic fields over the TI surface as follows. By using Eq. (\ref{BosonicOperators}), we determine the constant $\Lambda _{i}$ (appearing in the normalized amplitude) as a function of the bosonic operators. Therefore, in the steady-state condition, the electric field (\ref{ElectricField}) can be quantized as
\begin{align}
\vec{\mathcal{E}} (\vec{r}, t) & = \sqrt{\frac{\hbar \Omega}{2 \epsilon _{0} V _{m}}} \sum _{i = x,y,z} ( \hat{a} _{i} + \hat{a} _{i} ^{\dagger} ) \, \vec{\mathcal{Y}} _{i} (\vec{r}\, ) ,  \label{QuantizedEField}
\end{align}
where the mode volume $V _{m}$ is defined as the ratio between the total energy to the energy density inside the the TI $\mathcal{U} _{0}$, i.e.,
\begin{align}
V _{m} = \frac{1}{\mathcal{U} _{0}} \int \!\! d ^{3} \vec{r} \; \vert \vec{\mathcal{G}} _{i} (\vec{r} \, ) \vert ^{2}   \Bigg[ \frac{d [\mbox{\small Re}( \omega \epsilon ) ]}{d \omega} \Big| _{\omega = \Omega} + \frac{\mu _{\mbox{\scriptsize e}} ^{2} \tilde{\alpha} ^{2}}{4 \mu} \xi ^{2} (r) \Bigg] , \label{ModeVolumen}
\end{align}
where
\begin{align}
    \mathcal{U} _{0} = \vert \vec{\mathcal{G}} _{i} (\vec{0}\, ) \vert ^{2} \Bigg[ \frac{d ( \omega \epsilon _{1} )}{d \omega} \Big| _{\omega = \Omega} + \frac{\mu _{\mbox{\scriptsize e}} ^{2} \tilde{\alpha} ^{2}}{\mu _{1}} \Bigg] .  \label{EnergyDensity}
\end{align}
Further, $\vec{\mathcal{Y}} _{i} (\vec{r}\, ) = \vec{\mathcal{G}} _{i} (\vec{r}\, ) / \sqrt{\mathcal{U} _{0}}$ is a rescaled version of $\vec{\mathcal{G}} _{i} (\vec{r}\, )$. The total energy density inside the TI (\ref{EnergyDensity}) and the mode volume (\ref{ModeVolumen}) are explicitly evaluated in Appendix \ref{AppNormalization}.

\subsection{Hamiltonian of the system} \label{HamiltonianSection}

Having quantized the electromagnetic field operators, we can now define the Hamiltonian of the system. On the one hand, the Hamiltonian describing the topological insulator is given by Eq. (\ref{HamOperatorTI}), which corresponds to that of a mechanical oscillator. On the other hand, the part of the Hamiltonian describing the interaction between the TI and the dipole can be taken as $\hat{H} _{\mbox{\scriptsize int}} = - \, \hat{p} \cdot \vec{\mathcal{E}}$, where $\hat{p}$ is the dipole operator and $\vec{\mathcal{E}}$ is the quantized electric field operator given by Eq. (\ref{QuantizedEField}). When the distance between the dipole and the TI surface is larger than the TI radius, the dipolar approximation provides a reasonable estimate for the interaction. However, when the dipole is close enough to the TI surface, the effect of higher-order multipole moments becomes particularly important. See Refs. \cite{Anger, Yan} for experimental and theoretical studies regarding the validity of the dipolar approximation between quantum dots and metallic nanoparticles, which we assume to be feasible as well when the sample is a TI.

In general, the dipole operator can be expanded as $\hat{p} = \sum _{i,j} \vec{p} _{ij} \left| i \rangle \langle j \right|$, where $\vec{p} _{ij} = \langle i \vert \vec{p} \, \vert j \rangle $ are its matrix elements. For a spherically symmetric dipole we can choose a quantization direction such that $\vec{p} _{ij}$ points along an specific direction. Therefore, the interaction Hamiltonian will couple field operators (dipole and electric field) pointing along the same direction. So, if we excite only one specific transition of the dipole, we can make the two-level dipole approximation, and hence we can treat the TI-QD hybrid as a two-level system interacting with a single bosonic mode. Applying the two-level approximation \cite{ref2c2,ref2b2}, the dipole operator takes the form $\hat{p} = d (\hat{\sigma} _{+} + \hat{\sigma} _{-}) \hat{e} _{l}$ (with $l = x,y,z$), where $d$ is the dipole moment of the transition and $\hat{\sigma} _{+}$ and $\hat{\sigma} _{-}$ are the Pauli raising and lowering operators respectively. Dropping the index in the bosonic operators corresponding to the single TI mode interacting with the dipole, the interaction Hamiltonian can be written as
\begin{align}
    \hat{H} _{\mbox{\scriptsize int}} = \hbar g(r) (\hat{\sigma} _{+} + \hat{\sigma} _{-}) ( \hat{a} ^{\dagger} + \hat{a} ) , \label{InteractionHamiltonian}
\end{align}
where $g(r)$ is the TI-QD coupling strength. This coupling depends on whether, the electric field points along the dipole direction (longitudinal coupling) or in the transverse direction (transverse coupling). The explicit form of the coupling strength or Rabi frequency is then
\begin{align}
    g(r) = \left\lbrace \begin{array}{l}  +2 \, \frac{d}{\hbar} \sqrt{\frac{\hbar \Omega}{2 \epsilon _{0} V _{m} \mathcal{U} _{0}}} \frac{R ^{3}}{r ^{3}}   \\     -1 \, \frac{d}{\hbar} \sqrt{\frac{\hbar \Omega}{2 \epsilon _{0} V _{m} \mathcal{U} _{0}}} \frac{R ^{3}}{r ^{3}}   \end{array} \right. \;\;\; \begin{array}{l}      \mbox{longitudinal coupling}   \\      \mbox{transverse coupling}    \end{array} . \label{CouplingStrength}
\end{align}
This result implies that the Rabi frequency is twice as strong for longitudinal coupling, indicating that this is the preferable configuration for experimental detection of the TI effect. In this paper we will consider only the energy conserving terms in the interaction Hamiltonian (\ref{InteractionHamiltonian}), such that it reduces to $\hat{H} _{\mbox{\scriptsize int}} = \hbar g(r) (\hat{\sigma} _{+} \hat{a} + \hat{\sigma} _{-} \hat{a} ^{\dagger} )$.

The above analysis allows us to write the Hamiltonian of the closed system as
\begin{align}
    \hat{H} _{\mbox{\scriptsize S}} = \hat{H} _{\mbox{\scriptsize TI}} + \hat{H} _{\mbox{\scriptsize dip}} + \hat{H} _{\mbox{\scriptsize int}} ,
\end{align}
where $\hat{H} _{\mbox{\scriptsize TI}}$ is the Hamiltonian of the electromagnetic field excitations over the TI surface, given by Eq. (\ref{HamOperatorTI}), $\hat{H} _{\mbox{\scriptsize int}}$ is the interaction Hamiltonian of Eq. (\ref{InteractionHamiltonian}), and $\hat{H} _{\mbox{\scriptsize dip}} = \hbar \omega _{a} \hat{\sigma} _{+} \hat{\sigma} _{-}$ is the dipole Hamiltonian (being $\omega _{a}$ the resonant frequency of the dipole). Quantum dots subject to an electric field exhibit additional electro-absorption effects, including the quantum-confined Stark effect. This consists in small shifts in the energy levels, proportional to the squared magnitude of the electric field and the QD polarizability $f$, which can be properly ignored for electric fields satisfying $|\mathcal{\vec{E}}| \ll 2 d / f$. In Sec. \ref{NumericalSection} we estimate the numerical values for this effect and conclude that it can be safely disregarded in the present work.

In a realistic model, we have to include also damping effects due to the interaction of the system with the environment. To this end, here we consider that the system is coupled with a continuum reservoir of radiative output modes and a reservoir of phonon modes. The Hamiltonian of the radiative and phonon reservoirs is given by \cite{ref2c7,ref2c1}
\begin{align}
\hat{H} _{\mbox{\scriptsize B}} =\int \hbar \omega ^{\prime} ( \hat{b} ^{\dag} {}_{\!\! \omega'} \hat{b}_{\omega ^{\prime}} + \hat{c} ^{\dag} {}_{\!\! \omega'} \hat{c}_{\omega ^{\prime}} ) d\omega ^{\prime} , \label{ec99a}
\end{align}
while the Hamiltonian describing the interaction between the system and the reservoirs is
\begin{align}
\hat{H}_{\mbox{\scriptsize SB}} & = i\hbar\int ( T_{1}\hat{b} ^{\dag} {}_{\!\! \omega'} \hat{a} + T _{2} \hat{c} ^{\dag} {}_{\!\! \omega'} \hat{a} + T _{3} \hat{b} ^{\dag} {}_{\!\! \omega'} \hat{\sigma} _{-} ) d \omega ^{\prime} + \mbox{H.c.} \label{ec99b}
\end{align}
Here, $\hat{b} ^{\dag} {}_{\!\! \omega'}$ and $\hat{b} _{\omega'}$ ($\hat{c} ^{\dag} {}_{\!\! \omega'}$ and $\hat{c}_{\omega'}$) are the creation and annihilation operators corresponding to the radiative (phonon) modes. The terms $T_{1}=\sqrt{\gamma_{r}/2\pi}$ and $T_{2}=\sqrt{\gamma_{0}/2\pi}$ represent the coupling strength between the TI and the reservoir modes, while $T_{3}=\sqrt{\gamma_{s}/2\pi}$ represents the coupling strength between the dipole and the radiative modes. Further, $\gamma_{r}$, $\gamma_{0}$, and $\gamma_{s}$ are the scattering rate into free-space modes, the energy dissipation due to ohmic losses, and the spontaneous emission rate of the dipole, respectively. In Appendix \ref{GammaR}, by using the classical formulas for the electric and magnetic dipole radiation, we derive an exact expression for the scattering rate $\gamma _{r}$ into free-space modes for a TI. All in all, the full Hamiltonian of the open system can be written as
\begin{equation}
    \hat{H} = \hat{H} _{\mbox{\scriptsize TI}} + \hat{H} _{\mbox{\scriptsize dip}} + \hat{H} _{\mbox{\scriptsize int}} + \hat{H} _{\mbox{\scriptsize B}} + \hat{H}_{\mbox{\scriptsize SB}} . \label{FullHamiltonian}
\end{equation}

\section{Absorption spectrum} \label{AbsSpectrumSect}

The method of Zubarev's Green functions has found interesting  applications in different branches of physics. It was first conceived and successfully applied to different problems in statistical physics and linear response theory \cite{ref2d1,ref2d2,ref2d3,ref2d4}, and then it was adapted to study the optical absorption properties of hybrid systems formed by plasmonic nanoparticles and quantum emitters \cite{Manjavacas}. Outstandingly, it allows one to compute the absorption spectra from the retarded Zubarev-Green (ZG) function of the quantum operators that mediate the photon absorption process. After briefly recalling the basics of the method of Zubarev's Green functions, in this section we calculate the optical absorption spectrum of the topological insulator-quantum dot hybrid.

\subsection{Method of Zubarev's Green functions}

As is widely known, for a system initially in a state $\left| i \right>$ of energy $E _{i}$ that undergoes a transition to the final state $\left| f \right>$ of energy $E _{f}$, the optical absorption cross section is given by Fermi's golden rule according to
\begin{align}
    \sigma(\omega)\propto \sum _{f}
\vert \langle f;n-1 \vert \hat{H}^{\prime} \vert i;n \rangle \vert^{2} \, \delta(\omega _{fi} -  \omega ) , \label{AbsSpectrum}
\end{align}
where $\langle f;n-1 \vert \hat{H}^{\prime} \vert i;n \rangle$ is the matrix element of the perturbation $\hat{H}^{\prime}$ that couples the system with the external photon field, $\omega _{fi} = ( E _{f} - E _{i} ) / \hbar $ is the frequency corresponding to the difference between the final and initial energies, and $n$ is the number of external photons with frequency $\omega$.

In the problem at hand, the perturbation Hamiltonian has the generic form $\hat{H}^{\prime} \propto \hat{A} \hat{a} ^{\dagger} + \hat{A} ^{\dagger} \hat{a}$, where $\hat{a}$ and $\hat{A}$ ($\hat{a}^{\dagger}$ and $\hat{A}^{\dagger}$) are the annihilation (creation) operators for the external photons and excitations of the system, respectively. In this way, $\hat{A}$ connects the initial and final states of the system, and as such it governs the optical absorption properties. Equation (\ref{AbsSpectrum}) can be further simplified. Using the Sokhotsky's formula for the Dirac delta, $\delta (x) = \frac{1}{\pi} \mbox{Im}\{1/(x - i 0 ^{+})\}$, and the action of the bosonic ladder operators upon the basis $\left\{\ \!\! \left|n \right> \right\}\ $, we can recast Eq. (\ref{AbsSpectrum}) into
\begin{align}
    \sigma(\omega)\propto \mbox{Im} \sum _{f}
    \frac{\langle i|\hat{A}|f\rangle\langle f|\hat{A}^{\dagger}|i\rangle}
    {\omega _{fi}- \omega-i0^{+}} . \label{AbsSpectrum2}
\end{align}
Since we are considering a system in which external photons couple through excitation of a single quasiparticle excitation, we can safely assume that $\hat{A}^{\dagger}$ connects the initial ground state with a set of final states with a common energy $E _{f}$. Thus, the denominator gets out of the sum in Eq. (\ref{AbsSpectrum2}), and the expression for the cross section reduces to
\begin{align}
    \sigma(\omega) \propto \mbox{Im} \frac{\langle i|\hat{A}\hat{A}^{\dag}|i\rangle}{\omega _{fi}- \omega-i0^{+}} , \label{AbsSpectrum3}
\end{align}
where we have used the closure relation for the final states. Interestingly, the absorption spectrum is now written in terms of the expectation value of $\hat{A}\hat{A}^{\dag}$ in the initial ground state.

We now consider the definition of the retarded ZG function of two operators $\hat{A}$ and $\hat{B}$ in the frequency domain:
\begin{align}
    \langle \langle \hat{A} ; \hat{B} \rangle \rangle _{\omega+i0^{+}} = \frac{1}{i \hbar} \int _{0} ^{\infty} dt e ^{i(\omega + i 0 ^{+}) t} \theta (t) \langle [\hat{A} (t), \hat{B} (0)] _{\eta} \rangle , \label{GZ-function}
\end{align}
where $A (t)$ means the Heisenberg representation, $\theta (x)$ is the usual step function, and the brackets $[\hat{A}, \hat{B}] _{\eta} = \hat{A} \hat{B} - \eta \hat{B} \hat{A}$ stand for the commutator (anticommutator) of bosonic (fermionic) operators for $\eta = 1$ ($\eta = -1$). Now we take $\hat{B} = \hat{A} ^{\dagger}$, which is the appropriate choice to analyze the cross section (\ref{AbsSpectrum3}). For a fixed excitation frequency $\omega _{fi}$ the time-evolved (annihilation) excitation operator reads $\hat{A} (t) = \hat{A} (0) e ^{- i \omega _{fi} t }$, and hence, when the system is in the ground state, the retarded ZG function (\ref{GZ-function}) simplifies to
\begin{align}
    \langle \langle \hat{A} ; \hat{A} ^{\dagger} \rangle \rangle_{\omega + i 0^{+}} = - \frac{ \langle  \hat{A}(0) \hat{A} ^{\dag}(0) \rangle }{\omega _{fi}- \omega-i0^{+}} . \label{GZ-function2}
\end{align}
Finally, since $\langle i|\hat{A}\hat{A}^{\dag}|i\rangle$ and $\langle  \hat{A}(0) \hat{A} ^{\dag}(0) \rangle$ are both the same in the Schr\"{o}dinger and Heisenberg picture, respectively, the optical absorption cross section is then related to the retarded ZG function by
\begin{align}
     \sigma(\omega) \propto - \mbox{Im} \; \langle \langle \hat{A} ; \hat{A} ^{\dagger} \rangle \rangle _{\omega + i 0 ^{+}}  . \label{AbsSpectrum4} 
\end{align}
So, by computing the retarded ZG function $\langle \langle \hat{A} ; \hat{A} ^{\dagger} \rangle \rangle$ we will immediately obtain the optical absorption spectrum. In order to calculate $\langle \langle \hat{A} ; \hat{A} ^{\dagger} \rangle \rangle$, we shall use its equation of motion \cite{ref2d1}:
\begin{align}
    \hbar \omega \langle \langle \hat{A} ; \hat{A} ^{\dagger} \rangle \rangle = \langle[\hat{A} , \hat{A} ^{\dagger}] _{\eta} \rangle + \langle \langle [\hat{A} , \hat{H} ] ; \hat{A} ^{\dagger} \rangle \rangle  \label{EqMotion} ,
\end{align}
where $\hat{H}$ is the Hamiltonian of the system. Note that this expression depends on another ZG function $\langle \langle [\hat{A} , \hat{H} ] ; \hat{A} ^{\dagger} \rangle \rangle$, which can also be calculated by writing down its equation of motion. Iterating this process, one obtains a hierarchy of equations that may need to be truncated at some point by applying a physical approximation. This program will produce a linear system of equations from which we will obtain $\langle \langle \hat{A} ; \hat{A} ^{\dagger} \rangle \rangle$. In the next section we will compute the retarded ZG function $\langle \langle \hat{A} ; \hat{A} ^{\dagger} \rangle \rangle$ by using the Hamiltonian of Eq. (\ref{FullHamiltonian}).

\subsection{Optical absorption of the TI-QD hybrid}

In the problem at hand the bosonic operators $\hat{a}$ and $\hat{a} ^{\dagger}$ describe annihilation and creation of particle excitations on the TI surface. Hence, the optical absorption spectrum can be found from the retarded ZG function $\langle \langle \hat{a} ; \hat{a} ^{\dagger} \rangle \rangle$. The above equation of motion then reads
\begin{align}
    \hbar \omega \langle \langle \hat{a} ; \hat{a} ^{\dagger} \rangle \rangle = 1 + \langle \langle [ \hat{a} , \hat{H} ] ; \hat{a} ^{\dagger} \rangle \rangle , \label{EqMot1}
\end{align}
where $\hat{H}$ is the full Hamiltonian of the open system. Substituting the Hamiltonian (\ref{FullHamiltonian}) into Eq. (\ref{EqMot1}) we obtain
\begin{align}
   & \hbar ( \omega - \Omega ) \langle \langle \hat{a} ; \hat{a} ^{\dagger} \rangle \rangle = 1 + \hbar g (r) \, \langle \langle \hat{\sigma} _{-} ; \hat{a} ^{\dagger} \rangle \rangle \notag \\ & \hspace{1.4cm} - i \hbar \int [ T _{1} ^{\ast} \langle \langle \hat{b} _{\omega ^{\prime}} ; \hat{a} ^{\dagger} \rangle \rangle + T _{2} ^{\ast} \langle \langle \hat{c} _{\omega ^{\prime}} ; \hat{a} ^{\dagger} \rangle \rangle ] d \omega ^{\prime} . \label{EqMot2}
\end{align}
From this expression, it is clear that we need to compute three additional ZG functions: $\langle \langle \hat{\sigma} _{-} ; \hat{a} ^{\dagger} \rangle \rangle$, $\langle \langle \hat{b} _{\omega ^{\prime}} ; \hat{a} ^{\dagger} \rangle \rangle$, and $\langle \langle \hat{c} _{\omega ^{\prime}} ; \hat{a} ^{\dagger} \rangle \rangle$. These ZG functions can be obtained from their equations of motion. The last two of them can be easily obtained from their equations of motion:
\begin{align}
    \hbar (\omega - \omega ^{\prime}) \langle \langle \hat{b} _{\omega ^{\prime}} ; \hat{a} ^{\dagger} \rangle \rangle &= i \hbar [T _{1} \langle \langle \hat{a} ; \hat{a} ^{\dagger} \rangle \rangle + T _{3} \langle \langle \hat{\sigma} _{-} ; \hat{a} ^{\dagger} \rangle \rangle ] , \notag \\ \hbar (\omega - \omega ^{\prime}) \langle \langle \hat{c} _{\omega ^{\prime}} ; \hat{a} ^{\dagger} \rangle \rangle &= i \hbar T _{2} \langle \langle \hat{a} ; \hat{a} ^{\dagger} \rangle \rangle . \label{EqMot3}
\end{align}
Substituting these results into Eq. (\ref{EqMot2}) and solving for $\langle \langle \hat{a} ; \hat{a} ^{\dagger} \rangle \rangle$ we obtain
\begin{align}
   &  \langle \langle \hat{a} ; \hat{a} ^{\dagger} \rangle \rangle = \frac{1 + \hbar [ g (r) + \varpi _{13} ] \, \langle \langle \hat{\sigma} _{-} ; \hat{a} ^{\dagger} \rangle \rangle}{\hbar ( \omega - \Omega - \varpi _{11} - \varpi _{22} )}  , \label{EqMot4}
\end{align}
where we have defined the frequencies,
\begin{align}
    \varpi _{ij} &= \int \frac{T _{i} ^{\ast} (\omega ^{\prime}) T _{j} (\omega ^{\prime})}{\omega - \omega ^{\prime} - i 0 ^{+}} d \omega ^{\prime} \notag \\ &= \mbox{p.v.} \int \frac{T _{i} ^{\ast} (\omega ^{\prime}) T _{j} (\omega ^{\prime})}{\omega - \omega ^{\prime}} d \omega ^{\prime} + i \pi T _{i} ^{\ast} (\omega) T _{j} (\omega) , \label{NewFrequencies}
\end{align}
which have been evaluated using the Sokhatsky-Weierstrass theorem. Here, p.v. stands for the Cauchy principal value. Physically, the first term of $\varpi _{ij}$ represents a frequency shift, while the second one is a decay rate.

We still need to evaluate $\langle \langle \hat{\sigma} _{-} ; \hat{a} ^{\dagger} \rangle \rangle$. The corresponding equation of motion produces
\begin{align}
   & \hbar ( \omega - \omega _{a} ) \langle \langle \hat{\sigma} _{-} ; \hat{a} ^{\dagger} \rangle \rangle = \hbar g (r) \, \langle \langle ( 1 - 2 \hat{\sigma} _{+} \hat{\sigma} _{-} ) \hat{a} ; \hat{a} ^{\dagger} \rangle \rangle \notag \\ & \hspace{1.4cm}  - i \hbar \int T _{3} ^{\ast} \langle \langle ( 1 - 2 \hat{\sigma} _{+} \hat{\sigma} _{-} ) \hat{b} _{\omega ^{\prime}} ; \hat{a} ^{\dagger} \rangle \rangle d \omega ^{\prime} . \label{EqMot5}
\end{align}
Therefore, we see that new ZG functions emerge that need to be computed. The iteration process would produce an infinite number of equations of motions, but we truncate it at this point by approximating the operator $\hat{\sigma} _{+} \hat{\sigma} _{-}$ by its expectation value $\langle \hat{\sigma} _{+} \hat{\sigma} _{-} \rangle = n$. Taking this approximation and substituting Eq. (\ref{EqMot3}) into the above expression we obtain
\begin{align}
   & \langle \langle \hat{\sigma} _{-} ; \hat{a} ^{\dagger} \rangle \rangle = \frac{(1-2n) [ g (r) + \varpi _{31} ]}{\omega - \omega _{a} - (1-2n) \varpi _{33}} \, \langle \langle  \hat{a} ; \hat{a} ^{\dagger} \rangle \rangle . \label{EqMot6}
\end{align}
Finally, inserting these results into Eq. (\ref{EqMot4}), we obtain explicitly the ZG function $\langle \langle \hat{a} ; \hat{a} ^{\dagger} \rangle \rangle$ and hence the optical absorption spectrum of the TI-QD system:
\begin{align}
    \sigma (\omega) & \propto \mbox{Im} \; \left[ \omega - \mathfrak{W} - \frac{\Gamma}{\omega - {\mathcal{W}}} \right] ^{-1} , \label{AbsEspectrumFin}
\end{align}
where
\begin{align}
    \mathfrak{W} &= \Omega + \varpi _{11} + \varpi _{22} , \quad {\mathcal{W}} = \omega _{a} + (1-2n) \varpi _{33} , \notag \\ \Gamma &= (1-2n) [g(r) + \varpi _{13}] [g(r) + \varpi _{31}] .
\end{align}
This expression shows that the resonance frequency of the quantum modes on the TI is modified by the interaction with the quantum dot.

\subsection{Numerical results and discussion} \label{NumericalSection}

Here we apply our results to a realistic topological-insulator--quantum-dot hybrid, we have to use appropriate values for the different parameters appearing in Eq. (\ref{AbsEspectrumFin}). To this end, we consider the experimental setup used in Refs. \cite{ref31,ref32}. There the authors engineer a metallic-nanoparticle--quantum-dot hybrid by encapsulating gold nanospheres and cadmium selenide (CdSe) QDs in a polymer layer such as poly(methyl methacrylate). While it is difficult to perform this kind of experiments with topological insulators nowadays, we envision that a similar experimental setup could be engineered to test our results with a spherical topological insulator nanoparticles (as those predicted in Ref. \cite{STINano}) and CdSe quantum dots. It is worth mentioning that our results can also be tested with linear magnetoelectrics such as Cr$_{2}$O$_{3}$ \cite{CrO} or in some multiferroics \cite{Khomskii}. We will back to this discussion in the last section.

In order to calculate the optical absorption spectrum (\ref{AbsEspectrumFin}) we focus on the specific example of a CdSe quantum dot interacting with a TI spherical nanoparticle of TlBiSe$_2$. The values for the parameters which characterizes this TI has been experimentally investigated neglecting free carrier contributions and assuming high-frequency transparency \cite{Sato}. They are found to be $\mu _{1} = 1$ and $\epsilon _{1} (0) = 1 + (\omega _{e} / \omega _{R}) ^{2} \sim 4$ and have a single resonant frequency near 56 cm$^{-1}$ ($\sim 1.6$ THz) \cite{Mitsas}. We take an energy for the TI of $\hbar\Omega=2.2$ eV and the scattering rate into free-space modes $\gamma _{r}$, is obtained by substituting the above values into the formula (\ref{ScatteringRate}). Further, the damping parameter $\gamma _{0}$ satisfies $\gamma _{0} \ll \omega _{R}$, and hence it plays a secondary role. Therefore we can safely neglect it in the numerical simulations. The spherical TI nanoparticle is assumed to have a radius of $R \sim 5$ nm and be embedded in a nonmagnetic material with permittivity $\epsilon_{2} = 1.5$, which can be attained by capping the system with polymethyl methacrylate. We assume a CdSe quantum dot with a size of 4.5 nm; we take a spontaneous emission decay rate $\gamma _{s} ^{-1} = 10$ ns and a wavelength $\lambda=550$ nm, which is a common value for interactions with CdSe nanocrystals \cite{Waks}. The resonance energy $\hbar \omega _{a}$ is within the 1.5--2.9 eV range, an appropriate energy range for CdSe \cite{Manjavacas,ref32}. Also, for the transition dipole moment we set $d = 7.2\times10^{-28}$ Cm \cite{Eliseev,ref4c1}. As discussed in Sec. \ref{HamiltonianSection}, QDs subject to an applied electric field exhibit electro-absorption effects. Taking these numerical values into account, together with the polarizability of a CdSe quantum dot $f = 1.5 \times 10 ^{5} \textup{\r{A}}$ \cite{nanoscale2018}, we find that the energy shifts due to the Stark effect is of the order of $\sim 10^{-3}$eV, which is small as compared with the resonance energy so that we can safely disregard this effect in the present calculations. Finally, we suppose the system is under strong optical pumping, so that $n \approx 0.5$.

\begin{figure*}
\begin{center}
\includegraphics[scale=0.42]{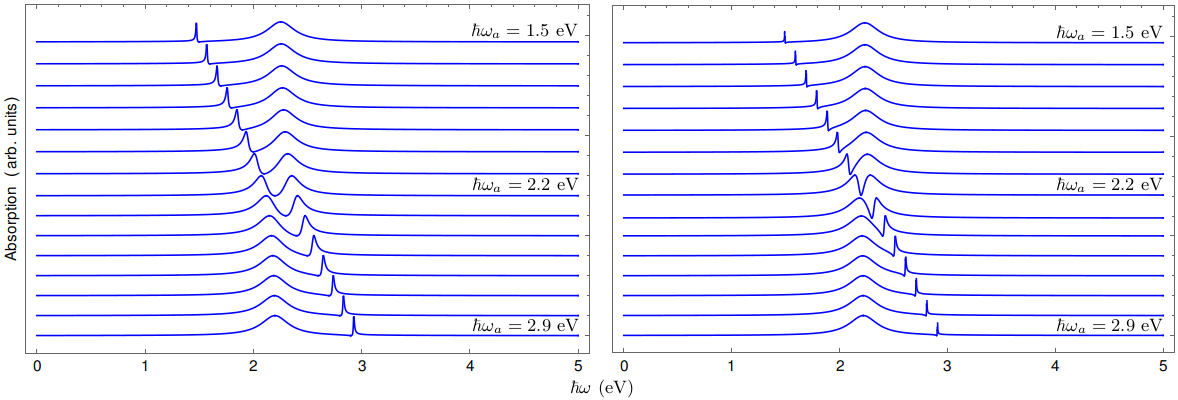}
\caption{Optical absorption spectrum of the TI-QD hybrid for longitudinal (left) and transverse (right) coupling and different values of QD resonance energy $\hbar\omega_{a}$. The TI-QD Fano resonance achieves its maximum approach for $\hbar\omega_{a}=2.2$ eV.}
\label{esp1}
\end{center}
\end{figure*}

\begin{figure*}
\begin{center}
\includegraphics[scale=0.48]{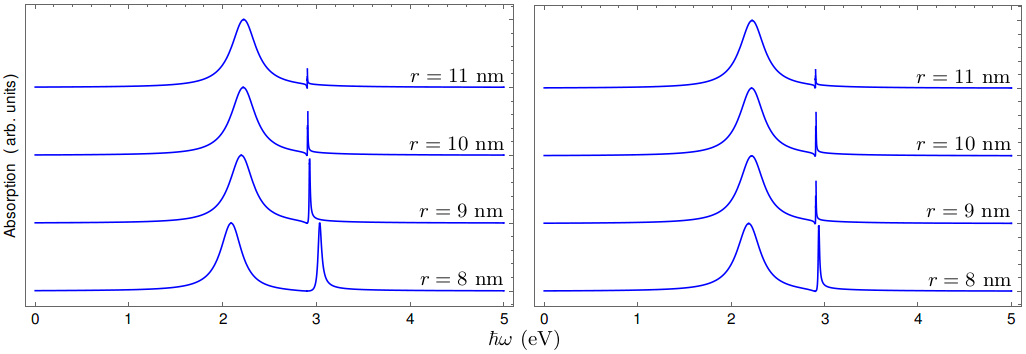}
\caption{Optical absorption spectrum of the TI-QD hybrid for longitudinal (left) and transverse (right) coupling for the distances $r = 8,\,9,\,10,\,11$ nm. We fix the QD resonance energy $\hbar\omega_{a}=2.9$ eV.}
\label{esp5}
\end{center}
\end{figure*}

Figure \ref{esp1} shows the optical absorption spectrum of the system under study, computed from Eq. (\ref{AbsEspectrumFin}), for longitudinal (at left) and transverse (at right) coupling. As we can see, the optical spectra exhibits interesting features. Outstandingly, a Fano resonance appears with a line shape that strongly depends on the energy $\hbar \omega _{a}$. The Fano resonance results from the interaction between a continuum of modes and a narrow discrete mode. In the problem at hand, the QD resonance is the narrow mode, while the bosonic excitations at the TI surface play the role of a continuum. Fano resonances in nontopological plexcitonic systems have been extensively studied, see for example \cite{Fano, Miroshnichenko, Farzad}. On the other hand, in fig. \ref{esp5} we show the optical absorption spectrum for different values of the QD-TI distance $r$, fixed $\hbar\omega_{a}=2.9$ eV and $\theta = \pi $. At left (right) we present the longitudinal (transverse) coupling. Here, we observe that QD-peak decreases as increasing $r$, implying that Fano resonances appears only when the QD is close to the TI surface. This is so because in this regime the interaction strength $g(r)$ increases as $r \to R$.

\begin{figure*}
\begin{center}
\includegraphics[scale=0.48]{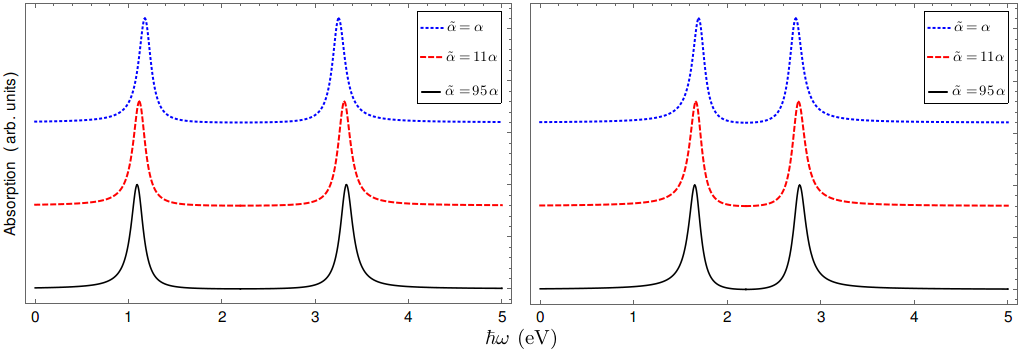}
\caption{Optical absorption spectrum of the TI-QD hybrid for longitudinal (left) and transverse (right) coupling for the rescaled topological magnetoelectric  polarizabilities $\tilde{\alpha}=\alpha$ (dotted blue line), $\tilde{\alpha}=11\alpha$ (dashed red line), and $\tilde{\alpha}=95\alpha$ (continuous black line). Here we fix $\hbar\omega_{a}=2.2$ eV and $r=7$ nm.}
\label{esp2}
\end{center}
\end{figure*}

We finally illustrate, in Fig. \ref{esp2}, the optical absorption spectrum as a function of the (rescaled) topological magnetoelectric polarizability $\tilde{\alpha}$, fixing $\hbar \omega _{a} = 2.2$ eV and $r = 7$ nm. Let us recall that the value of $\theta$ depends on the nature of the TR breaking perturbation at the TI surface. In practice, it corresponds to modifying the TI interface by adsorbing surface layers of nonzero Chern number. In Fig. \ref{esp2} we take $\tilde{\alpha} = \alpha$ (which is the lowest nontrivial possible value), $\tilde{\alpha} = 11 \alpha$ and $\tilde{\alpha} = 95 \alpha$. In general, we observe that the transverse coupling exhibits a distinctive feature: The two peaks approach each other faster than those appearing with the longitudinal coupling. In Fig. \ref{esp2} we have taken a large value of the magnetoelectric polarizability, which although it is very large for topological insulators, is appropriate for a magnetodielectric material such as Cr$_{2}$O$_{3}$, which is also described by axion electrodynamics. It would also be interesting to see the optical absorption spectrum when the QD interacts with an intrinsic magnetic topological insulator, such as the recently discovered MnBi$_2$Te$_4$ \cite{ExpMagneticTI, ExpMagneticTI2}. As discussed above, our analytical results are still valid for magnetic TIs. However, this compound was discovered the last year, and therefore we lack its full optical properties for the time being.

Recently, the authors in Ref. \cite{Farzad} introduced the concept of topological Fano resonance to name ultrasharp asymmetric line shapes which are protected against geometrical disorder of the sample, yet remain sensitive to external parameters. It was experimentally observed in acoustic systems \cite{Farzad}. In fact, it is worth mentioning that the Fano resonances we have obtained in the topological-insulator-quantum-dot hybrid, do not belong to the classification of topological Fano resonances of Ref. \cite{Farzad}. This is so because the topological protection of topological insulators is related to the band structure of the material, and not directly to its geometrical form. So, the Fano resonances we report are topological in a different sense.

\section{Summary and conclusions} \label{ConclusionSection}

The study and design of devices capable of controlling light-matter interaction at the nanoscale have been subjects of intense activity over the past decade. This interest has been reinforced by the recent advances in the fabrication of nanostructured devices made from topological insulator materials, such as TI nanoparticles and TI nanowires. Inspired by these studies, together with the lack of confirmation of the topological magnetoelectric effect of TIs, in this paper we have considered an hybrid system composed of a quantum-dot and a topological insulator nanoparticle, subject to a probe electric field. 

In order to study the optical response of this system we have employed a powerful quantum-mechanical approach which is commonly used to study the internal evolution (beyond the perturbative regime) of plasmonic nanostructures interacting with quantum emitters under strong optical pumping, and which takes into account energy loss due to spontaneous emission, ohmic losses, and scattering into free-space modes. The major advantage of this program is that it can be extended to include several kinds of interactions and energy loss mechanisms. The method relies on Zubarev's Green functions, which allow one to account for quantum aspects of the optical response and Fano resonances in plexcitonic systems. By using the above described method, we have expressed the optical absorption cross section in terms of the ZG function $\langle \langle \hat{a} ; \hat{a} ^{\dagger} \rangle \rangle _{\omega + i 0 ^{+}}$, where $\hat{a}$ and $\hat{a} ^{\dagger}$ are the bosonic operators describing creation and annihilation of quasiparticle excitations at the TI surface.

We applied our results to a realistic system which consists of a topological insulator nanoparticle (made of TIBiSe$_{2}$) interacting with a cadmium selenide quantum dot, both immersed in a polymer layer such as poly(methyl methacrylate). The optical absorption spectrum is found to exhibit Fano resonances resulting from the TI-QD interaction, similar to what happened with a metallic nanoparticle interacting with a quantum emitter \cite{Manjavacas,ref37,ref38}. As expected, transverse coupling and short TI-QD distances favor the absorption of the system, and this implies a better possibility to be experimentally detected. This conclusion about the strength of the topological magnetoelectric effect is similar to what occurs with the intensity of the monopole magnetic field appearing when a pointlike charge \cite{Qi-monopole} or a finite size sphere at constant potential \cite{MU2} is close to a planar TI surface. Further, we observe that high values of the magnetoelectric polarizability $\theta$ significantly shift the position of the absorption resonances. This suggests that this effect could also be explored with other magnetodielectric materials such as Cr$_{2}$O$_{3}$, which can be described by a similar axion coupling. Indeed, this material is characterized by a uniaxial magnetoelectric susceptibility tensor $\gamma _{ij}$, whose average is of the order of $\Bar{\gamma} \approx 0.7$ ps/m \cite{Hehl}. We leave this problem for a future work.

We close by commenting on possible extensions of this work. On the one hand, we can use the same idea to study the optical absorption of a quantum emitter placed in the gap of a topological insulator nanoparticle dimer. Additionally, the interaction between topological insulator quantum wires and quantum dots can also be analyzed within the same scheme. These problems will be further considered elsewhere.

\acknowledgements

A.M.-R. and L.C.-E. acknowledge support from DGAPA-UNAM project IA101320. L.F.Q. acknowledges support from SEP-CONACYT under project No. 288856.

\appendix

\section{SPHERICAL TI IN A CONSTANT ELECTRIC FIELD} \label{CalculationsFields}

In this section we present the detailed solution of the field equations for an spherical TI in a constant electric field, as shown in Fig. \ref{TI-QD-Fig}. Since there are no free-sources in the problem, one can introduce the electric $\phi$ and magnetic $\psi$ scalar potentials satisfying the Laplace equation
\begin{align}
\vec{\nabla} ^{2} \phi = 0 , \qquad \vec{\nabla} ^{2} \psi = 0 , \label{LaplacePotentials}
\end{align}
such that the electric and magnetic fields are given by $\vec{E} = - \vec{\nabla} \phi $ and $\vec{B} = - \vec{\nabla} \psi $, respectively. We choose coordinate axes such that the field $\vec{E} _{0}$ points along an arbitrary $u$-direction and the origin coincides with the center of the sphere, as shown in Fig. \ref{TI-QD-Fig}. Due to the axial symmetry of the problem (around the $u$-axis), the scalar potentials will be of the form $\phi = \phi (r , \varphi) $ and $\psi = \psi (r , \varphi) $, where $r$ is the radial coordinate and the polar angle $\varphi$ is defined through $\cos \varphi = u / r$. 

The general solutions of Eq. (\ref{LaplacePotentials}) for the potentials inside ($r<R$) and outside ($r>R$) the topological insulator can be expressed as an expansion in Legendre polynomials:
\begin{align}
\phi _{\mbox{\scriptsize in}} (r , \varphi) &= \sum _{\ell} A _{\ell} \, r ^{\ell} P _{\ell} (\cos \varphi) , \notag  \\ \phi _{\mbox{\scriptsize out}} (r , \varphi) &= - E _{0} r \cos \varphi + \sum _{\ell} \frac{C _{\ell}}{r ^{\ell +1}} P _{\ell} (\cos \varphi) , \notag \\ \psi _{\mbox{\scriptsize in}} (r , \varphi) &= \sum _{\ell} D _{\ell} \, r ^{\ell} P _{\ell} (\cos \varphi) , \notag \\ \psi _{\mbox{\scriptsize out}} (r , \varphi) &= \sum _{\ell} \frac{F _{\ell}}{r ^{\ell +1}} P _{\ell} (\cos \varphi) , \label{PotentialsExpansion}
\end{align}
where $r$ is the radial coordinate and the polar angle $\varphi$ is defined through $\cos \varphi = u/r$.

To solve the boundary value problem, we first note that, due to the asymptotic form of the electric field, $\vec{E} _{0} = E _{0} \vec{e} _{u}$, only the dipole terms with $\ell = 1$ will survive in the potentials (\ref{PotentialsExpansion}) once the boundary conditions (\ref{BoundaryConditions}) are imposed. The application of the boundary conditions (\ref{BoundaryConditions}) produces the following system of algebraic equations (with only $\ell = 1$ terms):
\begin{align}
\epsilon _{2} \left( 2C _{1} + E _{0} R ^{3} \right) + \epsilon _{1} A _{1} R ^{3} &= - \tilde{\alpha} c \, D _{1} R ^{3} , \notag \\ D _{1} R ^{3} &= - 2 F _{1} , \notag \\ R ^{3} \left( E _{0} + A _{1} \right) &= C _{1} , \notag \\ F _{1} / \mu _{2} - D _{1} R ^{3} / \mu _{1} &= - ( \tilde{\alpha} / c ) \, A _{1} R ^{3} ,
\end{align}
whose solution is quite simple. After straightforward calculations we obtain, for the electromagnetic fields inside the topological insulator,
\begin{align}
\vec{E} _{\mbox{\scriptsize in}} (\vec{r}, \omega) &= \frac{3 \epsilon _{2}}{2 \epsilon _{2} + \epsilon _{1} + \mu _{\mbox{\scriptsize e}} \tilde{\alpha} ^{2}} \vec{E} _{0} , \label{Ein} \\ \vec{B} _{\mbox{\scriptsize in}} (\vec{r}, \omega) &= \frac{3 \epsilon _{2} \mu _{\mbox{\scriptsize e}} \tilde{\alpha} / c}{2 \epsilon _{2} + \epsilon _{1} + \mu _{\mbox{\scriptsize e}} \tilde{\alpha} ^{2}} \vec{E} _{0} , \label{Bin}
\end{align}
where $\mu _{\mbox{\scriptsize e}} = 2 \mu _{1} \mu _{2} / (\mu _{1} + 2 \mu _{2})$. An interesting feature to note is the form of the fields in such region. The electric field behaves as the field produced by a uniformly polarized sphere, while the magnetic field resembles the one produced by a uniformly magnetized sphere.

Outside the TI, the electromagnetic fields read
\begin{align}
\vec{E} _{\mbox{\scriptsize out}} (\vec{r}, \omega) &= \vec{E} _{0} + \frac{1}{4 \pi \epsilon _{2}} \left[ \frac{3 (\vec{p} \cdot \vec{r}) \vec{r}}{r ^{5}} - \frac{\vec{p}}{r ^{3}} \right] , \label{Eout} \\ \vec{B} _{\mbox{\scriptsize out}} (\vec{r}, \omega) &= \frac{\mu _{2}}{4 \pi} \left[ \frac{3 (\vec{m} \cdot \vec{r}) \vec{r}}{r ^{5}} - \frac{\vec{m}}{r ^{3}} \right] , \label{Bout}
\end{align}
where 
\begin{align}
\vec{p} \, (\omega) &= 4 \pi \epsilon _{2} \frac{\epsilon _{1} - \epsilon _{2} + \mu _{\mbox{\scriptsize e}} \tilde{\alpha} ^{2}}{2 \epsilon _{2} + \epsilon _{1} + \mu _{\mbox{\scriptsize e}} \tilde{\alpha} ^{2}} R ^{3} \vec{E} _{0} , \\ \vec{m} (\omega) &= \frac{2 \pi}{\mu _{2}} \frac{ 3 \epsilon _{2} \mu _{\mbox{\scriptsize e}} \tilde{\alpha} / c}{2 \epsilon _{2} + \epsilon _{1} + \mu _{\mbox{\scriptsize e}} \tilde{\alpha} ^{2}} R ^{3} \vec{E} _{0} . 
\end{align}
The electric field consists of a superposition of the applied electric field and that of a pointlike electric dipole $\vec{p} \, (\omega)$ located at the origin and pointing in the direction of the applied field. The magnetic field can be interpreted as that generated by a magnetic dipole $\vec{m} (\omega)$ located also at the origin. In the limit $\tilde{\alpha} \to 0$, the fields above reduce to the well-known results in the literature.

One can further check that Eqs. (\ref{Ein}) and (\ref{Eout}) for the electric field can be written in a unified fashion as $\vec{E}_{0} + \vec{\mathcal{E}} (\vec{r} , \omega)$, where $\vec{\mathcal{E}} (\vec{r} , \omega)$ is given by Eq. (\ref{E-TI}). Also, Eqs. (\ref{Bin}) and (\ref{Bout}) for the magnetic field can be written as in Eq. (\ref{B-TI}).

\section{TI RESPONSE WITH $\omega$-DEPENDENT PERMITTIVITY} \label{Approximations}

When substituting the dielectric function (\ref{DielectricFunction}) into Eqs. (\ref{E-TI}) and (\ref{B-TI}) we obtain the expressions
\begin{align}
\vec{\mathcal{E}} (\vec{r}, \omega) &= \sum _{i} \frac{[\omega (\omega + i \gamma _{0}) - \omega _{R} ^{2}] \frac{\epsilon _{2} - 1 - \mu _{\mbox{\scriptsize e}} \tilde{\alpha} ^{2}}{2 \epsilon _{2} + 1 + \mu _{\mbox{\scriptsize e}} \tilde{\alpha} ^{2}} + \omega _{0} ^{2}}{\Omega ^{2} - \omega (\omega + i \gamma _{0})} E _{0i} \, \vec{\mathcal{G}} _{i} (\vec{r}) , \label{E-app} \\ \vec{\mathcal{B}} (\vec{r}, \omega) &= \sum _{i} (\eta \, \mu _{\mbox{\scriptsize e}} \, \tilde{\alpha}  / 2 c) \, \xi (r) \frac{\omega _{R} ^{2} - \omega (\omega + i \gamma _{0})}{\Omega ^{2} - \omega (\omega + i \gamma _{0})} E _{0 i} \, \vec{\mathcal{G}} _{i} (\vec{r}) , \label{B-app}
\end{align}
where $\omega _{0} = \omega _{e} / \sqrt{2 \epsilon _{2} + 1 + \mu _{\mbox{\scriptsize e}} \tilde{\alpha} ^{2}}$, $\eta = 3 \epsilon _{2} / (2 \epsilon _{2} + 1 + \mu _{\mbox{\scriptsize e}} \tilde{\alpha} ^{2})$ and $\Omega ^{2} = \omega _{0} ^{2} + \omega _{R} ^{2}$. We consider the case where $\gamma _{0} \ll \omega$. In this limit, the denominator of the above expressions can be written as
\begin{align}
\Omega ^{2} - \omega (\omega + i \gamma _{0}) \approx - 2 \omega [(\omega - \Omega) + i \gamma _{0} / 2] ,
\end{align}
which is the standard approximation for a high-$Q$ resonator. In this limit the TI response will be very small unless $\omega \approx \Omega$. Using the assumption that $\omega \approx \Omega \gg \gamma _{0}$ the numerator in Eq. (\ref{E-app}) can be simplified to
\begin{align}
    [\omega (\omega + i \gamma _{0}) - \omega _{R} ^{2}] \frac{\epsilon _{2} - 1 - \mu _{\mbox{\scriptsize e}} \tilde{\alpha} ^{2}}{2 \epsilon _{2} + 1 + \mu _{\mbox{\scriptsize e}} \tilde{\alpha} ^{2}} + \omega _{0} ^{2} \approx \eta \, \omega _{0} ^{2} 
\end{align}
while the numerator in Eq. (\ref{B-app}) simplifies to
\begin{align}
    \omega _{R} ^{2} - \omega (\omega + i \gamma _{0}) \approx - \omega _{0} ^{2} .
\end{align}
Inserting these approximations into Eqs. (\ref{E-app}) and (\ref{B-app}) we attain Eqs. (\ref{E-TI2}) and (\ref{B-TI2}).

\section{EVALUATION OF THE NORMALIZATION FACTOR AND MODE VOLUME} \label{AppNormalization}

From the definition of the normalization factor in Eq. (\ref{Normalization}) we find that
\begin{align}
    \frac{2 \hbar \Omega}{\epsilon _{0} \mathcal{N} ^{2}} \! &= \! \left( \! \epsilon _{2} + \frac{\mu _{\mbox{\scriptsize e}} ^{2} \tilde{\alpha} ^{2}}{4 \mu _{2}} \right) \! I _{1} + \! \Bigg\{\ \!\! \frac{d [ \mbox{Re}( \omega \epsilon _{1} ) ]}{d \omega} \Bigg| _{\omega = \Omega}  \!\!\!\! + \! \frac{\mu _{\mbox{\scriptsize e}} ^{2} \tilde{\alpha} ^{2}}{\mu _{1}} \Bigg\}\ \!\! I _{-1} , \label{NormalizationEv}
\end{align}
where we have defined the integral
\begin{align}
I _{s} = \int d ^{3} \vec{r} \; \vert \vec{\mathcal{G}} _{i} (\vec{r} \, ) \vert ^{2} \; \Theta \left[ s (r - R) \right] .
\end{align}
Here, $\Theta [x]$ is the Heaviside step function. These integrals can be easily evaluated. Using the form of the vector $\vec{\mathcal{G}} _{i} (\vec{r} \, )$, given by Eq. (\ref{G-vector}), we obtain
\begin{align}
    I _{-1} &= \! \int \!\! \Theta (R - r) d ^{3} \vec{r} = \! \int _{0} ^{2 \pi} \!\!\! \int _{0} ^{\pi} \!\!\! \int _{0} ^{R} \!\! r ^{2} \sin \theta dr d \theta d \varphi = \frac{4 \pi R ^{3}}{3} , \notag \\ I _{+1} &= \! \int \!\! \Theta (r - R) \frac{1 + 3 ( \vec{e} _{i} \cdot \vec{e} _{r} ) ^{2}}{(r/R) ^{6}}  d ^{3} \vec{r} \notag \\ &= \! \int _{0} ^{2 \pi} \!\!\! \int _{0} ^{\pi} \!\!\! \int _{R} ^{\infty} \! \frac{1 + 3 \cos ^{2} \theta}{(r/R) ^{6}} r ^{2} \sin \theta dr d \theta d \varphi = \frac{8 \pi R ^{3}}{3} . \label{Integrals}
\end{align}
On the other hand, the derivative in the second term of the right-hand-side in Eq. (\ref{NormalizationEv}) can be evaluated with the help of the dielectric function model (\ref{DielectricFunction}). In the limit $\Omega \gg \gamma _{0}$ we obtain
\begin{align}
    \frac{d [ \mbox{Re}( \omega \epsilon _{1} ) ]}{d \omega} \Bigg| _{\omega = \Omega} \!\!\!\! = 1 + (\omega _{e} / \omega _{0}) ^{2} + 2 (\omega _{e} / \omega _{0}) ^{4} (\omega _{R} / \omega _{e}) ^{2} .
\end{align}
By using these results, together with the definition for the ratio $\omega _{0} / \omega _{e}$ , Eq. (\ref{NormalizationEv}) yields
\begin{align}
    \frac{1}{\mathcal{N} ^{2}} &= \frac{4 \pi \epsilon _{0} R ^{3}}{3 \hbar \Omega} \frac{(2 \epsilon _{2} + 1 + \mu _{\mbox{\scriptsize e}} \tilde{\alpha} ^{2} ) \big[ 2 \epsilon _{2} + \epsilon _{1} (0) + \mu _{\mbox{\scriptsize e}} \tilde{\alpha} ^{2} \big]}{\epsilon _{1} (0) - 1} ,
\end{align}
where $\epsilon _{1} (0) - 1 = (\omega _{e} / \omega _{R})^{2}$ is the static permittivity of the topological insulator.

In a similar fashion, one can further evaluate the energy density inside the TI, defined by Eq. (\ref{EnergyDensity}). The result is
\begin{align}
    \mathcal{U} _{0} = 2 (\epsilon _{2} + 1) + \mu _{\mbox{\scriptsize e}} \tilde{\alpha} ^{2} (1 + \mu _{\mbox{\scriptsize e}} / \mu _{1}) + 2 \frac{(2 \epsilon _{2} + 1 + \mu _{\mbox{\scriptsize e}} \tilde{\alpha} ^{2} )}{\epsilon _{1} (0) - 1} ,
\end{align}
and with the help of the above results we obtain that the mode volume, defined by Eq. (\ref{ModeVolumen}), is
\begin{align}
    V _{m} = \frac{8 \pi R ^{3}}{3} \frac{(2 \epsilon _{2} + 1 + \mu _{\mbox{\scriptsize e}} \tilde{\alpha} ^{2} ) \big[ 2 \epsilon _{2} + \epsilon _{1} (0) + \mu _{\mbox{\scriptsize e}} \tilde{\alpha} ^{2} \big]}{\mathcal{U} _{0} [\epsilon _{1} (0) - 1]} .
\end{align}

\section{RADIATIVE DECAY OF THE TI} \label{GammaR}

When the TI is excited by an input electric field of the form $\vec{E} _{0} (\vec{r} , t) = E _{0} \delta (t) \vec{e} _{z}$, the electric and magnetic fields due to the TI at time $t>0$ are given by $\vec{\mathcal{E}} (\vec{r} ,t) = \Lambda \sin ^{2} (\Omega t ) \vec{\mathcal{G}} _{z} (\vec{r})$ and $\vec{\mathcal{B}} (\vec{r},t) = (\mu _{\mbox{\scriptsize e}} \tilde{\alpha} / 2 c) \xi (r) \vec{\mathcal{E}} (\vec{r},t)$, respectively. The functions $\vec{\mathcal{G}} _{i} (\vec{r})$ and $\xi (r)$ are defined in the Eq. (\ref{G-vector}). Further, from Eq. (\ref{EnergyStored2}) we find that the energy stored in the electromagnetic fields, averaged over an optical cycle, is given by
\begin{align}
\left< U \right> =  \frac{1}{2} \hbar \Omega \, (\Lambda / \mathcal{N}) ^{2} , \label{AverageEnery}
\end{align}
where $\Lambda = E _{0} \eta (\omega _{0} ^{2} / 2 \Omega)$ and $\mathcal{N}$ is the normalization factor computed in Appendix \ref{AppNormalization}. Outside the TI, the electromagnetic fields take the form of dipole fields, with electric dipole moment $\vec{p} = 4 \pi \epsilon _{0} \epsilon _{2} \Lambda R ^{3} \vec{e} _{z}$ and magnetic dipole moment $\vec{m} = (\mu _{\mbox{\scriptsize e}} \tilde{\alpha} c/2 \epsilon _{2} \mu _{2})\vec{p}$ oscillating at the frequency $\Omega$. So, in terms of the electric dipole moment $p$, the averaged energy (\ref{AverageEnery}) can be written as
\begin{align}
    \left< U \right> =  \frac{1}{2} \hbar \Omega \, \frac{\vec{p} ^{\, 2}}{(4 \pi \epsilon _{0} \epsilon _{2} R ^{3} \mathcal{N}) ^{2}} . \label{EnergyAveraged}
\end{align}
If the TI radius is much smaller than the wavelength of the input field, we can treat the TI as a pair of point dipoles. Therefore, the Larmor formula is appropriate to compute the power radiated by the dipoles, i.e., $\mathcal{P} = (\mu _{0} \mu _{2}/12 \pi c)(\Ddot{\vec{p}} ^{\, 2} + \frac{1}{c ^{2}} \Ddot{\vec{m}} ^{2} )$. Substituting the above values for the dipole moments we obtain 
\begin{align}
    \mathcal{P} = - \frac{d \left< U \right>}{dt} = \frac{\mu _{0} \mu _{2} p ^{2} \Omega ^{4}}{12 \pi c} \left[ 1 + \left( \frac{\mu _{\mbox{\scriptsize e}} \tilde{\alpha}}{2 \epsilon _{2} \mu _{2}} \right) ^{2} \right] , \label{RadiatedPower}
\end{align}
and the expressions (\ref{EnergyAveraged}) and (\ref{RadiatedPower}) are related by $\mathcal{P} = \gamma _{r} \left< U \right>$, from which we obtain the decay rate of the TI due to scattering into free-space modes,
\begin{align}
    \gamma _{r} = \frac{2 \mu _{2} \epsilon _{2} ^{2} R ^{3} \Omega ^{4}}{c ^{3}} \frac{[\epsilon _{1} (0) - 1] \big[ 1 + (\mu _{\mbox{\scriptsize e}} \tilde{\alpha} / 2 \epsilon _{2} \mu _{2}) ^{2} \big]}{(2 \epsilon _{2} + 1 + \mu _{\mbox{\scriptsize e}} \tilde{\alpha} ^{2} ) \big[ 2 \epsilon _{2} + \epsilon _{1} (0) + \mu _{\mbox{\scriptsize e}} \tilde{\alpha} ^{2} \big]} . \label{ScatteringRate}
\end{align}

\end{document}